\documentclass[conference]{IEEEtran}
\usepackage{graphicx}
\usepackage{algorithmic}
\usepackage{array}
\usepackage[colorlinks=true,citecolor=blue,linkcolor=blue,urlcolor=blue,plainpages=false]{hyperref}
\usepackage{booktabs, tabularx, multirow, array}
\usepackage{breakurl}
\usepackage{nomencl}
\usepackage{multirow}
\usepackage{comment} 
\usepackage{amsmath}
\usepackage{amsfonts}
\usepackage{amssymb}

\def\x{\mathbf{x}}
\def\z{\mathbf{z}}
\def\e{\mathbf{e}}
\def\H{\mathbf{H}}
\def\a{\mathbf{a}}
\def\c{\mathbf{c}}
\def\W{\mathbf{W}}
\def\r{\mathbf{r}}

\begin{document}

\title{Cyberattack Detection in Intelligent Grids \\
Using Non-linear Filtering}

\author{\IEEEauthorblockN{Irina Lukicheva \\ David Pozo }
\IEEEauthorblockA{Center for Energy Systems \\
Skolkovo Institute of Science and Technology \\
Moscow, Russia  \\
irina.lukicheva@skolkovotech.ru; d.pozo@skoltech.ru}
\and
\IEEEauthorblockN{Alexander Kulikov}
\IEEEauthorblockA{Institute of Power Engineering \\
Nizhny Novgorod State Technical University n.a. Alekseev \\
Nizhny Novgorod, Russia \\
inventor61@mail.ru 
}
}

\maketitle

\begin{abstract}
Electric power grids are evolving towards intellectualization such as Smart Grids or active-adaptive networks.  Intelligent power network implies usage of sensors, smart meters, electronic devices and sophisticated communication network. This leads to a strong dependence on information and communication networking that are prone to threats of cyberattacks, which challenges power system reliability and efficiency. Thus, significant attention should be paid to the Smart Grids security. 

Recently, it has been proven that False Data Injection Attacks (FDIA) could corrupt results of State Estimation (SE) without noticing, therefore, leading to a possible mis-operation of the whole power system. 

In this paper, we introduce an algorithm for detecting cyberattacks based on non-linear filtering by using cyber-physical information from Kirchhoff laws. The proposed algorithm only needs data from adjacent nodes, therefore can be locally and distributed implemented. Also, it requires very low computational effort so that it can be run online, and it is suitable for implementation in existing or new ad-hoc low-cost devices. 

The proposed algorithm could be helpful to increase power system awareness against FDIA complementing the current SE implementations. The efficiency of the proposed algorithm has been proved by mathematical simulations and computer modeling in PSCAD software. Our results show that the proposed methodology can detect cyberattacks to the SE in 99.9\% of the cases with very little false alarms on the identification of spoiled measurements (4.6\%).  
\end{abstract}


\IEEEpeerreviewmaketitle

\section{Introduction}

Nowadays, electrical grids are moving towards intellectualization and automation resulting into cyber-physical systems. It implies extensive usage of smart meters, sensors and sophisticated communication infrastructures.  This provides many benefits for electrical grid, but at the same time, it makes power system more vulnerable to cyberattacks. A significant threat to cyber-physical systems is the possibility of physical damage to equipment or potential loss of the complete system. However, a significant advantage in power systems when dealing with cyber threats is that cyber-physical systems are subject to physical laws (Kirchhoff's laws). Thus, we could take advantage of exploiting information about the system’s inherent physical laws for developing new monitoring algorithms for cyberattack detection. 

\subsection{Motivation and Literature Review}
Among all possible cyber threats, we can highlight State Estimation (SE) cyberattacks \cite{he2016cyber}. State Estimation is a core procedure of control centers which processes raw measurements and provides reliable state of the power system. Although SE was originally introduced to deal with sensor measurement errors for finding data consistency \cite{schweppe1970power}, it was able to detect intentionally corrupted data in many cases. Generally, SE is very sensitive to measurements with gross error, which can corrupt the SE results. Bad data detection (BDD) is used to eliminate corrupted data. BDD is performed by residual-based $\chi^2$--test and largest normalized residual test. However, Liu et al. \cite{liu2011false}, introduced a False Data Injection Attack (FDIA) algorithm which is able to bypass BDD and maliciously change SE outcome that can lead to power system mis-operation and eventually its blackout.

Since the first introduction of the FDIA algorithm in \cite{liu2011false}, many authors have proposed several methods for power system defense against FDIA. They can be divided into two categories: a \textit{protection-based defense} and a \textit{detection-based defense} \cite{yang2014false}.  The first category includes methods aimed to secure selected measurement meters to not give an opportunity to launch a FDIA \cite{bi2014graphical,deng2017defending,bobba2010detecting,bhattarai2012novel}. The second category addresses methods can be referred to detection-based defense \cite{kim2011strategic,chaojun2015detecting,karimipourfalse,moslemi2017fast,zhou2016cyber,li2015quickest,zhu2016defending, jiang2017kalman}. 

In \cite{chaojun2015detecting}, authors compare the distribution of measurement variation from the historical data and the distribution of measurement variation between the current time step and the previous time step by calculating Absolute distance and Kullback--Leibler distance. An Euclidean detector was proposed in \cite{zhou2016cyber}. It can detect the distance deviation between measurements vector and estimated measurements vector. If the difference between the measurements and the estimated data is larger than a threshold, the detector will trigger an alarm. In \cite{karimipourfalse}, after each SE process, all states are checked for belonging to a Markov chain. If the estimated state is out of the Markov chain, there is a high possibility of being under a cyberattack. In \cite{moslemi2017fast}, Markov graph of the phase angles is used. Cumulative sum type algorithms based on the generalized likelihood ratio for centralized and distributed cyberattack detection were described in \cite{li2015quickest}. In \cite{zhu2016defending}, a detection algorithm is proposed based on FDIA stealthness corruption. The corruption is performed by replacing at least one of the suspect state variables with the reliable ones derived from historical running state database. When FDIA is not stealth anymore, it can be detected. In \cite{jiang2017kalman} a fast-$\chi^2$ detector is introduced supported by a diffusion Kalman filter, in which after every meter updates its present state that meter obtains a further optimal state by communicating with its neighbors. The proposed algorithm in \cite{ashok2016online} detects gross measurement anomalies due to stealthy cyberattacks by predicting states independently, using real-time short-term load forecast information, generation schedule information and real-time data from existing PMU deployments, and comparing them with state estimator outputs.
 
\subsection{Paper Contributions and Organization}

All methods mentioned above require either additional information (forecasted or historical) or performing SE before FDIA detection. In this paper, we propose a methodology for FDIA detection based on median filtering that does not require any knowledge of current power system state, measurement variance distribution before and/or after an attack, and any other historical data older than a one-time or three-time steps. Additionally, the proposed methodology is easy to implement and can be computed in distributed way (it only need information of adjacent nodes). Finally, the proposed methodology requires low computational effort so that it can be run online for fast FDIA detections.  

The rest of this paper is organized as follows. Section \ref{sec2} introduces the FDIA in the SE problem. Section  \ref{sec3} presents the median filtering technique for reconstructing measurements. Section  \ref{sec4} proposes the methodology for cyberattack detection. Section  \ref{sec5} illustrates the applicability of the proposed methodology. Finally, Section  \ref{sec6} some relevant concluding remarks are provided.

\section{False Data Injection Attack against State Estimation}
\label{sec2}

In \cite{liu2011false}, a FDIA class of cyberattacks was introduced and applied to the SE problem, demonstrating that it is possible to corrupt the SE without noticing, i.e, passing the $\chi^2$-test for BDD. The idea of this cyberattack is to inject malicious measurements that cannot be identified by BDD and spoil the result of state estimation. This approach is based on the principle that all BDD techniques rely on the assumption that the residuals (squares of differences between the measurements and their corresponding estimates) are large in case of bad data. However, there is a possibility to introduce coordinated malicious measurements simultaneously in several buses so that the assumption is breached, i.e. that the attack is not detected. 

The basic principle of FDIA is the following. Let assume that we want to estimate the state of a system defined by $n$ state variables (nodal voltage phasors, i.e., magnitude and phase angle voltages) defined by the vector $\x = [x_1, \ldots, x_n]^{\top}$. We assume that $m$ meters provide $m$ measurements defined by the vector $\z  = [z_1, \ldots, z_m]^{\top}$. In general, we can describe the measurements as a function of the state variables of the system, \eqref{eq1},
		
\begin{IEEEeqnarray}{lr}   
\z= \H (\x) + \e   \label{eq1} 
\end{IEEEeqnarray} 
		
where $\e$ is the vector of dimension $m \times 1$ of errors associated with the measurements $\z$ with normal distribution; $\H(\x)$ represents the non-linear Kirchhoff equations that relates states (complex voltage phasors) with measurements (active and reactive power, voltages, currents, etc.,) \cite{gomez2004power}. In our paper, measurements are obtained from PMUs (current and voltage phasors). Then, $\H(\x)$ would represent the Ohm's law only, therefore relation between states and measurements would be linear without any other approach\footnote{SE is in general non-linear for a conventional measurements from SCADA systems such as active power, but exist linear approximations for dealing with it \cite{gomez2004power}. Most of the results obtained in this paper can be applied to the linearized version of SE considering SCADA measurements without structural changes in the algorithms.}, i.e.,  $\H(\x) = \H \x$ . 

When $\H$ has rank equal to $n$, we can find the best estimator for the state variable, $\hat{\x}$, by minimizing the weighted square residuals \eqref{WLEM}, \cite{gomez2004power}. This method is the so-called Weighted Least Square Method (WLSM). 

\begin{IEEEeqnarray}{lr}   
\hat{\x} = \min_{\x} \W \,|| \z - \H\x||^2   \label{WLEM} 
\end{IEEEeqnarray}
 
For this particular problem \eqref{WLEM}, with linear dependency, exist analytical solution given by \eqref{Eq.analyticalSol}\footnote{since $\H$ is a complex matrix, under $\H^{\top}$ we understand conjugate-transpose matrix operation.}. The symbol $\W$ stands for the weight of every measurement. It is usually directly related with the accuracy (quality) of the measurement (meter).

\begin{IEEEeqnarray}{lr}   
\hat{\x} =  (\H^{\top}\W \H)^{-1}\H^{\top}\W \, \z   \label{Eq.analyticalSol} 
\end{IEEEeqnarray}

Now, assume that there is a measurement vector $\z_a$, which includes malicious data introduced by an attacker given by $\z_a = \z  + \a$, where $\z$  is original measurement vector and $\a  = [a_1, \ldots, a_m]^{\top}$ is the malicious attack vector. In general, the vector $\a$ should be detected as a bad data by using $\chi^2$-test analysis. However, in {\cite[Theorem 1]{liu2011false}} it was demonstrated that exist infinite number of attacks $\a$ that can bypass the $\chi^2$-test. In particular, undetectable attacks can be constructed by $\a = \H \c$, where $\c = [c_1, \ldots, c_n]^{\top}$ is any random vector that represents the state estimation error. Thus, the state estimated under a FDIA is given by $\hat{\x}_a = \hat{\x} + \c$. 

It is easy to demonstrate (see \eqref{eq.resi}) that the attack $\a$ is undetectable by showing that the residuals from the system under attack, $\r_a$, are the same that the residuals of the system without attack, $\r$. 

\begin{IEEEeqnarray}{rl}   
	\r_a = \z_a - \H \hat{\x}_a = \z + \a - \H (\hat{\x} + \c) & \nonumber \\ 
         = (\z - \H \hat{\x}) + (\a - \H \c)  &= \r   \label{eq.resi}
\end{IEEEeqnarray}

We should remark that the attack can aim to spoil certain states from the system that have high impact on the grid, with the possibility of triggering alarms or automatic disconnections of elements of the grid and a possible series of cascade outages and in the worst scenario causing a blackout. Also, it is possible to generate an attack vector without the need to spoil all measurements. Thus, if the $rank(H) = n$, it is needed to spoil one measurement for every state variable. Therefore, it is important to complement the conventional monitoring algorithms to increase the state of awareness. 

In the next section, we present a decentralized and computationally inexpensive algorithm for increasing the awareness at different levels.

\section{Median Filtering in Power Systems}
\label{sec3}

Median filtering (MF) is a nonlinear signal processing technique widely used in signal and image processing \cite{huang1981two}. Median filtering is performed by letting a window move over the points of a sequence and replacing the value at the window center with the median of the original values within window \cite{huang1981two}. The median of a $n$--odd sequence is the middle element of the sequence, when the sequence has been shorted ascendantly. 

\begin{figure}
	\centering
	\includegraphics[width=0.95\linewidth]{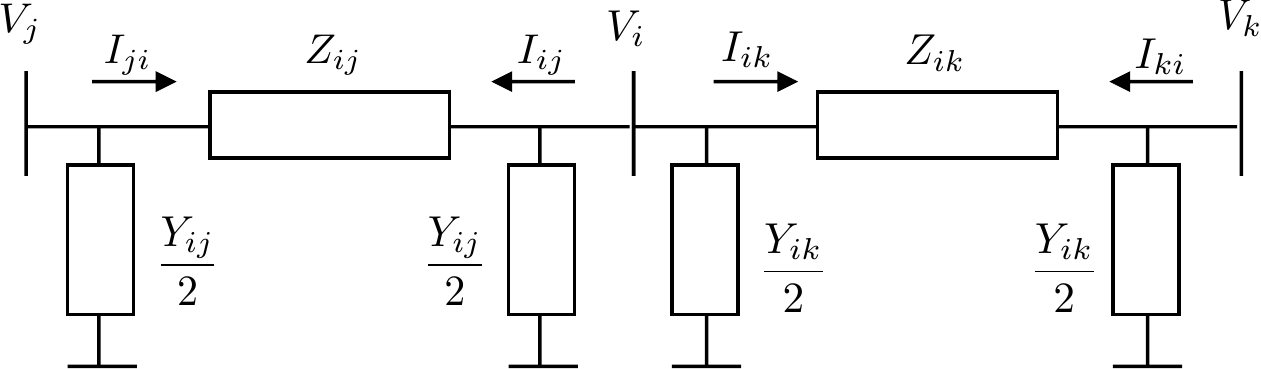}
	\caption{Three-bus example with a $\pi$--equivalent transmission line}
		\label{fig1}
\end{figure}

In power systems, it is possible to reconstruct electrical values in one bus using information measured from other buses due to the physical laws (Kirchhoff's laws) that couple one bus with another. In Figure \ref{fig1}, we show an example of three interconnected buses. In this case, the voltage state variable at node $i$, $V_i$ can be either measured at the node $i$, or reconstructed from node $j$ and/or $k$ by using Ohm's laws. The fact that power systems has to follow to physical laws allows us to obtain extra information about the state variables using data from other nodes.

As well as in image processing \cite{huang1981two}, in the application of cyberattack detection, one-dimensional and two-dimensional median filtering can be used.

\subsection{One-dimensional Median Filtering}

One-dimensional median filtering assumes only information coming from the same time frame or snapshot. For the sake of simplicity, we limit our approach to three values from adjacent buses to estimate the voltage state. Note that we could extend it to more than three values. Using the above example from Figure \ref{fig1}, we can define the voltage phasor estimation, $\hat{V}_i$, at node $i$ as

\begin{IEEEeqnarray}{lr}   
	\hat{V}_i  = \mathrm{median} (V_{i(i)} , V_{i(j)}, V_{i(k)} ), \label{eq4} 
\end{IEEEeqnarray} 
		
where $V_{i(i)} = V_i$ is the direct voltage phasor measurement at node $i$,  and $V_{i(j)}$, $V_{i(k)}$  are the voltage phasor measurement at the nodes $i$ obtained through Omh's laws, \eqref{eq5}, \eqref{eq6}, using  the measurements from nodes $j$ and $k$, respectively. 

\begin{IEEEeqnarray}{lr}   
	V_{i(j)} = \left( V_j  \frac{Y_{ij}}{2} - I_{ji}\right)   Z_{ij}   \label{eq5}   \\
	V_{i(k)} = \left( V_k  \frac{Y_{ik}}{2} - I_{ki}\right)   Z_{ik}   \label{eq6} 
\end{IEEEeqnarray}

\subsection{Two-dimensional Median Filtering}

Additionally to the spatial dimension, the temporal dimension can be used for reconstructing states values. 
Thus, for two-dimensional median filtering we have an square window of $3 \times 3$ defined in \eqref{eq7} for reconstructing voltage at node $i$.
		
\begin{IEEEeqnarray}{lr}   
		\hat{V}_i  = \mathrm{median}  \left(\begin{array}{ccc}
		V_{i(i)}(t_1)& V_{i(i)}(t_2)  &  V_{i(i)}(t_3) \\ 
        V_{i(j)} (t_1 )& V_{i(j)} (t_2) & V_{i(j)}(t_3)  \\ 
	    V_{i(k)} (t_1 )& V_{i(k)} (t_2) & V_{i(k)}(t_3)  \\ 
	\end{array} \right)  \label{eq7}
\end{IEEEeqnarray} 
		
where  $V_{i(i)}(t_1)$, $V_{i(i)}(t_2)$, and $V_{i(i)}(t_3)$ are the direct voltage measurements at node $i$  at the times  $t_1$, $t_2$, and $t_3$ respectively. Similarly to the one-dimensional case, we can obtain voltage phasor estimation from buses $k$ and $j$ at the times $t_1$, $t_2$, and $t_3$.

\section{Algorithm for FDIA Detection}
\label{sec4}

In this section, we make use of the previous section's results where Omh's law is used to generate new estimates of given measurements. It should be noted that the knowledge degree of the grid parameters will influence the results of the proposed algorithm. However, it was assumed that the parameters are well-known since they should be defined before the state estimation process. In addition, parameters uncertainty is a separate problem which is not considered in the paper.  

Several verification criteria for the detecting the presence of cyberattacks are listed below:
\begin{itemize}
\item	Voltage anomaly criteria;
\item	Direct current imbalance criteria;
\item	Calculated current imbalance criteria.
\end{itemize}

\subsection{Voltage Anomaly Criteria}
\label{subseV}

Let us conventionally call the set of values used in MF at every bus by a sequence  of elements $V_{i(n)}$, where the subscript $i$ denotes the node which this value refers to, and $n$ is the element number in the sequence. The elements of the sequence can be obtained directly from measurements or can be calculated using data from adjacent nodes. Then, we define the \textit{anomaly coefficient of voltages} for the node $i$ as the maximum of the deviation rates \eqref{eq. anomaly V}. The deviation rates are calculated with reference to the median value at every node $i$. Observe, that \eqref{eq. anomaly V} can be applied either to one- or two-dimensional median filtering.

\begin{IEEEeqnarray}{lr}   
	\kappa_i^V   =  \max \left(\frac{|V_{i(n)}- \hat{V}_i|}{\hat{V}_i}, \forall n \right)  \label{eq. anomaly V}
\end{IEEEeqnarray} 
		
\subsection{Direct Current Imbalance and Calculated Current Imbalances criteria}
\label{subseI}

We can define the \textit{direct current imbalance}, $\kappa_i^I $ , at node $i$ as the sum of all incoming and outgoing currents to and from the node \eqref{eq. anomaly I}. Observe that, due to first Kirchhoff's law this value should be zero. 

\begin{IEEEeqnarray}{lr}   
	\kappa_i^I   =  \sum_{n}  I_{in}  \label{eq. anomaly I}
\end{IEEEeqnarray} 
		
%where $n$ is the number of currents incoming to or outgoing from node $i$. 
Note that $I_{in}$ is the branch current phasor measurement from PMUs.
However, the incoming and outgoing currents phasors at node $i$ can be also estimated from the other interconnected nodes by

\begin{IEEEeqnarray}{lr}   
	\hat{I}_{in} = \left(\frac{V_{i} - V_n}{Z_{in}} \right) + V_i \frac{Y_{in}}{2}.  \label{eq. anomaly II}
\end{IEEEeqnarray} 

Similarly, the \textit{calculated current imbalance} $\kappa_i^{\hat{I}}$ is determined  by 
\begin{IEEEeqnarray}{lr}   
	\kappa_i^{\hat{I}}   =  \sum_{n}  \hat{I}_{in}.  \label{eq. anomaly I_cal}
\end{IEEEeqnarray}
		
%where n is the number of currents incoming to or outgoing from node i (Figure \ref{fig1}).

It was assumed that if any coefficient $\kappa_i^V$, $\kappa_i^I$, or $\kappa_i^{\hat{I}}$, is larger than the chosen threshold, it can signalize about the attack. 

\subsection{Illustrative Example}

\begin{figure}[hp]
	\centering
	\includegraphics[width=0.95\linewidth]{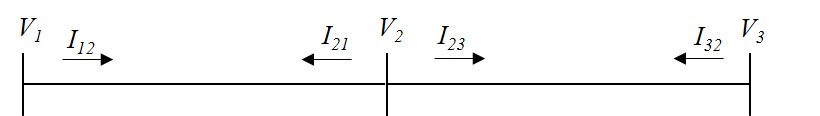}
	\caption{Illustrative example of a partial part of a power grid}
		\label{fig2}
\end{figure}

Let assume a part of an electrical grid shown in \mbox{Figure \ref{fig2}}.  There are measuring devices installed in nodes 1, 2, 3 and in lines 1--2 and 2--3. They provide measurements values $V_1$, $V_2$, $V_3$, $I_{12}$, $I_{21}$, $I_{23}$, $I_{32}$. The topology and all line parameters are known. 

The one-dimensional MF sequence for node 2 consists of direct measurement $V_2$; calculated voltage in node 2, using data from node 1, $V_{2(1)}$; and calculated voltage in node 2, using data from node 3, $V_{2(3)}$, that is, $\{V_2,V_{2(1)},V_{2(3)}\}$.

Let assume that the voltage measurement at node 2 was corrupted and we have got the measure \mbox{$V_2 = 1.5$ pu} while the other nodes used for calculations of $V_{2(1)}$,  and $V_{2(3)}$ are not corrupted, in this case, \mbox{$V_{2(1)}=1$ pu} and \mbox{$V_{2(3)}=1$ pu.} Then, the estimated voltage at node 2 is \mbox{$\hat{V_2}= \mathrm{median}(V_2,V_{2(1)},V_{2(3)}) = \mathrm{median}(1.5,1,1) = 1$ pu}.  Also, we calculate voltage anomaly coefficient of the MF sequence as follows.
		
 \begin{IEEEeqnarray}{rl}   
	\kappa_2^V & = \max \left(\frac{|V_{2}-\hat{V_2}|}{\hat{V_2}},\frac{|V_{2(1)}-\hat{V_2}|}{\hat{V_2}}, \frac{|V_{2(3)}-\hat{V_2}|}{\hat{V_2}} \right) \nonumber \\
    & = \max \left(0.5,0,0 \right)= 0.5 \label{eq13} 
\end{IEEEeqnarray} 
		
Withing a threshold of $\kappa_2^V>0.05$, we observe that the first measurement (voltage at node 2) is suspect of being corrupted. We mark it as an anomaly measurement.

\section{Numerical Example}
\label{sec5}

\begin{figure}[h]
	\centering
	\includegraphics[width=0.95\linewidth]{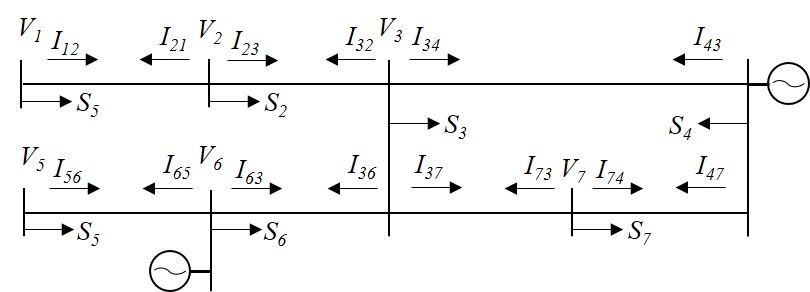}
	\caption{7-bus grid model}
		\label{fig3}
\end{figure}

For testing the performance of the proposed algorithm in this paper, we use a 7-bus electrical grid illustrated in \mbox{Figure \ref{fig3}}. 

A PSCAD model was used to obtain the measurement data of the proposed electrical grid. Since the number of PMU installed in power systems is constantly increasing, it was assumed that PMU are deployed on every bus, so all complex voltage and current vectors are measured. SCADA measurements for cyberattack detection could be used as well with minor changes, but this case is not considered in this paper.

Simulated measurement data was obtained from the PSCAD models for three sequential time periods and it was posteriorly processed in MATLAB, where, we implemented the algorithm presented in the previous sections. In MATLAB, at moment $t_3$, a false data injection attack was simulated following the algorithm described in Section \ref{sec2}. For one-dimensional MF, we took values from the same time period where the attack is generated, $t_3$. For two-dimensional MF algorithm, measurements at times $t_1, t_2, t_3$  were used. SE was performed by WLSM using corrupted measurements. The results of SE were compared with initial, ``clear'' from any errors, voltage values. 

We settled a threshold for attack detection equal to 5\%. That is if the difference between estimated state from MF and direct measurement or calculated state was larger than 5\%, we consider our electrical grid under attack. After that, all coefficients $\kappa_i^V$, $\kappa_{i}^I$, $\kappa_{i}^{\hat{I}}$  for every node were calculated to identify which measurement has an anomalous value. 

We generated 1000 Monte Carlo simulations. At every simulation a random attack vector to voltage and current measurements based on the methodology proposed in Section \ref{sec2} is created. The compromised voltage and current measures are choosing randomly for every simulation. As result, 1423 voltage measurements were under attack out of 7000 (7 nodes times 1000 random attacks).

\subsection{Results}
The result of different cyberattack simulations are presented in Table \ref{tab1} and Table \ref{tab4}. 
Column 2 and 3 from Tables \ref{tab1} to \ref{tab4} represent the percentage of cases where the FDIA was accurately detected (or undetected) at every particular node under attack using the anomaly voltage criteria presented at subsection \ref{subseV}. Column 2 refers to the one-dimensional MF (1D MF) while column 3 stands for the two-dimensional MF case (2D MF). Column 4 (DCI), stands for the direct current indicator \eqref{eq. anomaly I}, and column 5 (CCI), refers to the calculated current indicator \eqref{eq. anomaly I_cal}.

\begin{table}[h!] 
  \begin{center}
    \caption{Percentages of Detected and Not Detected Attacks in relation to the Total Number of Attacked Voltage Measurements}
    \label{tab1}
    \begin{tabular}{|c|c|c|c|c|} 
      \hline
      \textbf{Criteria} & \textbf{1D MF} & \textbf{2D MF} & \textbf{DCI} & \textbf{CCI}\\
      \hline
      Detected attacks, \% & 55.94 & 100 & 82.29 & 86.16\\
       \hline
      Not detected attacks, \% & 44.06 & 0 & 17.71 & 13.84\\
       \hline
    \end{tabular}
  \end{center}
\end{table}

\begin{table}[h!]
  \begin{center}
    \caption{Percentage of False Alarms and Detected absence of Attacks in relation to the Total Number of Not Attacked Voltage Measurements}
    \label{tab2}
    \begin{tabular}{|c|c|c|c|c|} 
      \hline
      \textbf{Criteria} & \textbf{1D MF} & \textbf{2D MF} & \textbf{DCI} & \textbf{CCI}\\
      \hline
      Detected absence of attacks \% & 78.90 & 78.88 & 100 & 74.90\\
       \hline
      False alarms, \% & 21.10 & 21.12 & 0 & 25.10\\
       \hline 
    \end{tabular}
  \end{center}
\end{table}

The performance of the proposed method differs from node to node. Tables \ref{tab1} and \ref{tab2} presented and aggregated view for all nodes of the system. However, there are some nodes that may be more difficult to detect when they are under attack. Thus, we have computed the percentages of detection (or not) for every node independently. The range of those percentage are given in the Table \ref{tab3} and Table \ref{tab4}. For example, we observe from Table \ref{tab3} and the CCI criteria that we are able to detect FDIA in some nodes 100\% of the attacks, while for other nodes under FDIA, we are able to detect them only 76.23\% of the time. When the system is under attack but a particular node is not, we found for the same criteria CCI, that there are nodes with false alarms ranging from 8.45\% to 53.39\%. 

\begin{table}[h!]
  \begin{center}
    \caption{Range  of Detected and Not Detected Attacks in relation to the Total Number of Attacked Voltage Measurements}
    \label{tab3}
    \begin{tabular}{|c|c|c|c|c|} 
      \hline
      \textbf{Criteria} & \textbf{1D MF} & \textbf{2D MF} & \textbf{DCI} & \textbf{CCI}\\
      \hline
      Detected   & & & & \\
      attacks, \% & 1.80--96.32 & 100 & 72.38--99.10 & 76.23--100\\
       \hline
      Not detected  & & & & \\
      attacks, \% & 3.68--98.20 & 0 & 0.90--26.72 & 0--23.75\\
       \hline
      
    \end{tabular}
  \end{center}
\end{table}

\begin{table}[h!]
  \begin{center}
    \caption{Range of False Alarms and  Detected absence of Attacks in relation to the Total Number of Not Attacked Voltage Measurements}
    \label{tab4}
    \begin{tabular}{|c|c|c|c|c|} 
      \hline
      \textbf{Criteria} & \textbf{1D MF} & \textbf{2D MF} & \textbf{DCI} & \textbf{CCI}\\
      \hline
      Detected absence & & & & \\
      of attacks \% & 78.91--90.2 & 78.91--90.2 & 100 & 46.61--91.55\\
       \hline
      False  alarms, \% & 9.8--36.54 & 9.8--36.54 & 0 & 8.45--53.39\\
       \hline      
    \end{tabular}
  \end{center}
\end{table}

From results shown in Tables \ref{tab1}--\ref{tab4} it can be observed that two-dimensional MF method is able to detects
attacks in 100\% cases. Nevertheless, it suffers from large value of false alarms. This problem is brought about the fact that large value of criterion coefficient can be caused not only by the measurement in the particular node which median sequence is considered, but also by any other measurement that participate in calculation of MF sequence elements. To decrease the number of false alarms, a new layer is added to the algorithm. It is introduced in the next subsection.  

It should be noted, that under false alarm we refer to the situation when the system is under attack, but the particular measurement input marked as corrupted is mistakenly selected. That is, the algorithm trigger an alarm by identifying a node as attacked while it is not.

\subsection{Algorithm for Reducing False Alarms}
\label{subsec5a}

To decrease the number of false alarms, we depict a conditional casuistic algorithm to discard measurement that was classified as under attack (voltage anomaly was over the given threshold) by the MF algorithm mentioned above. This new algorithm is iteratively applied to each node that was classified as under attack. We make use of the measurements that were not was not marked as suspect. We consider this measure not spoiled by an FDIA, but may be prone
to have an error associated to its own PMU measurement.

For instance, let us consider the MF sequence ($V_{1}$, $V_{1(2)}$, $V_{1(3)}$) for the information used to compute the state of $V_1$, marked as under attack. If the algorithm for FDIA detection defined $V_{1}$ and $V_{2}$ as a suspect but $V_{3}$ was not suspected, we consider $V_{3}$, therefore $V_{1(3)}$, as a trusted reference. Then, we can decide whether $V_{1}$, was really corrupted or it is a false alarm. That is if $\frac{|V_{1}-V_{1(3)}|}{\hat{V}_1} \leq \epsilon$, then $V_{1}$ is the true measurement, and it was a false alarm. Otherwise, $V_{1}$ will remain tagged as under attack. Observe that, if a node tagged under attack has all the adjacent nodes marked under attack it is not possible to apply new this algorithm. However, this procedure can be performed in several iterations, improving the accuracy of FDIA detection at every step. Note that, at every iteration, new nodes are reclassified from under  attack to normal condition. 

Summing up, the proposed cyberattack detection consists of two stages. In stage 1, suspected measurements are detected using voltage anomaly criteria. In stage 2, a refinement algorithm is performed for every node labeled as under attack. Thus, in stage 2 we specify whether the detected measurement is really corrupted or it was misclassified (false alarm).

The results of the algorithm for two-dimensional MF are presented in Tables \ref{tab5}. As we see, the detection of cyberattacks is very high with a very low number of false alarms. Additionally, we can observe in the second row that there is a relatively small dispersion in detecting at which bus is producing the cyberattack. Thus, the proposed tool results in an relative accurate algorithm for not only detecting that a network is compromised by a cyberattack, but also by localizing the measurements attacked.

\begin{table}[h!]
  \begin{center}
    \caption{Performance of Improved Two-dimentional MF Algorithm for Cyberattack Detection}
    \label{tab5}
    \begin{tabular}{|c|c|c|c|c|} 
      \hline
      \textbf{} & \textbf{Detected} & \textbf{Not detected} & \textbf{Detected absence} & \textbf{False}\\
      \textbf{} & \textbf{attacks} & \textbf{attacks} & \textbf{of attacks} & \textbf{alarms}\\
      \hline
      For all & & & & \\
      nodes, \% & 99.93 & 0.07 & 95.41 & 4.59\\
       \hline
      Range, \% & 99.8--100 & 0--0.2 & 90.63--98.99 & 1.01--9.37\\
       \hline   
    \end{tabular}
  \end{center}
\end{table}

\section{Conclusion}
\label{sec6}

In this paper an algorithm for cyberattack detection has been proposed. From the modeling results, it can be concluded that MF showed good performance. One of the most important advantages of MF is that it is easy to implement, it requires very low computational effort, and it can be distributively performed. 
The proposed algorithm it could be a useful tool, cheap and easy to implement, for fast cybersecurity surveillance and real-time situational awareness.

Finally, future directions of research are the application to distribution networks with intermittent renewable generation. Also, it is worth to explore median filtering models from a probabilistic perspective that it could complement the presented algorithm in this paper.

\section*{Acknowledgment}
This work was supported by Skoltech NGP Program (Skoltech-MIT joint project).

\bibliographystyle{IEEEtran}
\bibliography{references.bib}

\end{document}